\documentclass[usegraphicx,usenatbib]{mn2e}
\newcommand{\Msun}{\hbox{$\rm\thinspace M_{\odot}$}}

\begin{document}

\title[Quiescent variability from Aql X-1]{Quiescent X-ray variability from the
neutron star transient Aql X-1}

\author[Cackett et~al.]{E. M. Cackett$^1$\thanks{ecackett@ast.cam.ac.uk},
J. K. Fridriksson$^2$,
J. Homan$^2$,
J. M. Miller$^3$,
R. Wijnands$^4$
\\ $^1$ Institute of Astronomy, University of Cambridge, Madingley Rd,
Cambridge, CB3 0HA, UK
\\ $^2$ MIT Kavli Institute for Astrophysics and Space Research, 70 Vassar
Street, Cambridge, MA 02139, USA
\\ $^3$ Department of Astronomy, University of Michigan, 500 Church St, Ann
Arbor, MI 48109-1042, USA
\\ $^4$ Astronomical Institute `Anton Pannekoek',
        University of Amsterdam, Science Park 904, 1098 XH Amsterdam, the
Netherlands
	}
\date{Received ; in original form }
\maketitle

\begin{abstract}
A number of studies have revealed variability from neutron star low-mass X-ray
binaries during quiescence.  Such variability is not well characterised, or
understood, but may be a common property that has been missed due to lack of
multiple observations.  One such source where variability has been observed
is Aql X-1.  Here, we analyse 14 {\it Chandra} and {\it XMM-Newton} observations
of Aql~X-1 in quiescence, covering
a period of approximately 2 years. There is clear variability between the
epochs, with the most striking feature being a flare-like increase in the flux
by a factor of 5. Spectral fitting is inconclusive as to whether the power-law
and/or thermal component is variable. We suggest that the variability and
flare-like behaviour during quiescence is due to accretion at low rates which
might reach the neutron star surface.
\end{abstract}


\section{Introduction}

Transient neutron star low-mass X-ray binaries spend the majority of their
lifetime in a quiescent state and only a small amount of the time in outburst,
accreting at a significantly higher rate ($\sim 0.1 -- 1.0  L_{\rm edd}$). During quiescence
the X-ray spectrum is typically characterised by a thermal and/or power-law
component.  The thermal
component is most frequently interpreted as emission from the neutron
star surface.  In the deep crustal heating scenario \citep*{BBR98}, pycnonuclear
reactions occur in the inner crust during outburst due to compression of the
crust by accreted material.  Energy deposited by these reactions heats the
neutron star core on a timescale of $10^4 - 10^5$ yr, where it reaches a steady
state luminosity set by the time-averaged mass accretion rate.  The neutron star
should therefore have a minimum thermal luminosity due to these processes.  This
thermal emission is fit well by neutron star atmosphere models, which could open
the possibility of measuring the neutron star radius
\citep[e.g.,][]{rutledgeetal99}.

The origin of the power-law component, however, remains poorly understood.  A
number of emission mechanisms have been suggested involving on-going
low levels of accretion, but whether the gas impacts the neutron star surface
\citep[e.g.,][]{menoumcclintock01}, is stopped at the magnetospheric radius, or
whether there is a shock between a radio pulsar relativistic wind and matter
transferred from the companion star \citep[see the review by][]{campanaetal98a}
is unclear.  Moreover, on-going accretion onto the neutron star surface could
potentially produce a thermal-like spectrum \citep{zampierietal95}.

Variability has been observed during quiescence in a number of sources, with
variability seen over a wide range of timescales from hundreds of
seconds through to years
\citep{campana97,campanaetal04,rutledgeetal00,rutledgeetal01a,rutledge02,
cackett05,cackett10,muno07,fridriksson10,fridriksson11}. In the majority of
cases, the variability has either been
attributed to the power-law component or its origin remains unclear. However,
our recent observations of Cen~X-4 showed variability as large as a factor of
4.4 which could only be fit by variability in both the thermal and power-law
components \citep{cackett10}. Another source showing particularly interesting
quiescent variability is XTE~J1701$-$462 \citep{fridriksson10,fridriksson11}. 
In this source, an overall cooling of the thermal component has been observed
after a long outburst lasting $\sim$2 years. However, on two occasions a sudden
short-term flare has been observed, with the brightest one rising to a flux
about 20 times the normal quiescent level.  Such flares are presumably due to
sporadic increases in accretion rate during quiescence.

The focus of this current work is Aql X-1, one of the sources where
variability has been observed both on short
(hundreds of seconds) and long (months) timescales \citep{rutledge02}. 
However, the nature of this variability has been debated.  While
\citet{rutledge02} conclude that the thermal component is variable, a
subsequent analysis by \citet{campanastella03} showed that correlated changes
in the power-law index and column density can also describe the data.  Having
recently found large amplitude quiescent variability in Cen X-4, we also
searched for further variability in archival observations of Aql X-1.  In
addition to the 4 observations analysed by \citet{rutledge02} and
\citet{campanastella03}, we analyse a further 10 quiescent observations
performed about a year later. In section~\ref{sec:data} we detail the data
reduction, section~\ref{sec:spec} describes the spectral analysis, and in
section~\ref{sec:disc} we discuss our findings.

\section{Data Reduction}\label{sec:data}

\begin{figure}
\centering
\includegraphics[width=8.4cm]{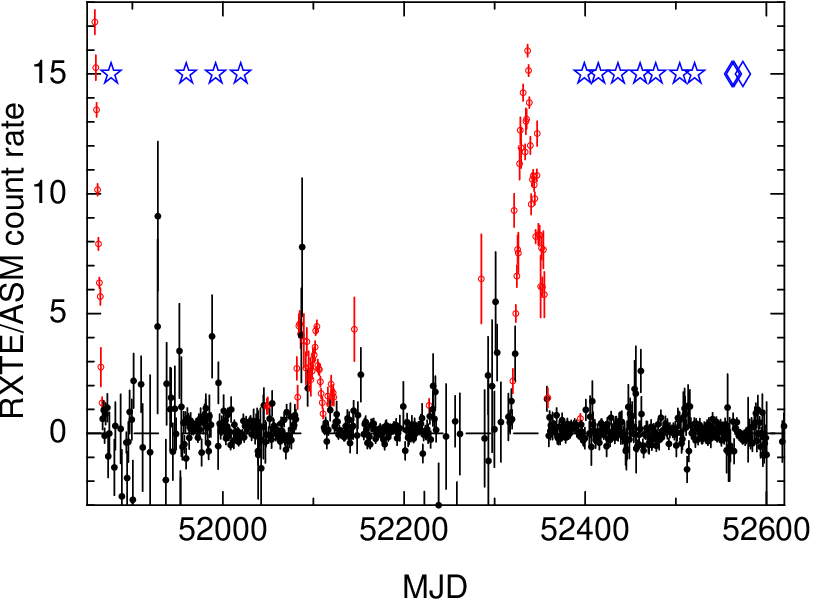}
\caption{RXTE/ASM 1-day averaged lightcurve of Aql X-1 around the times of the
{\it Chandra} and {\it XMM-Newton} quiescent observations.  Significant
detections (3$\sigma$) are marked as red open circles, whereas non-detections
are black filled circles.  The blue stars mark the times of the quiescent
observations with {\it Chandra} and blue diamonds mark the times of {\it
XMM-Newton} observations. }
\label{fig:asmlc}
\end{figure}

Here we analyse a total of 14 observations of Aql X-1 in a quiescent state (11
with {\it Chandra}, 3 with {\it XMM-Newton}), all the available
archival {\it Chandra} and {\it XMM-Newton} observations of Aql X-1 in
quiescence. In Table~\ref{tab:obs} we give
details of these 14 observations.  Figure~\ref{fig:asmlc} shows the one-day
averaged lightcurve from the {\it RXTE} All-sky monitor (ASM), with the times of
these observations marked. Two outbursts occur between the first four {\it
Chandra} observations and the last 10 {\it Chandra}/{\it XMM-Newton}
observations.  All observations analysed here are clearly at times when Aql X-1
is undetected by the {\it RXTE}/ASM.  The 1-day average detection threshold (2 -- 10 keV) for the ASM is approximately 10 mCrab \citep{levine96}.

\begin{table*}
\centering
\caption{Quiescent observations of Aql X-1}
\label{tab:obs}
\begin{tabular}{ccccccc}
\hline
Observation & Start date & MJD & Mission/Instrument & ObsID & Exp. time & Net count rate  \\
 & dd/mm/yy & mid observation & & & (ksec) & (c s$^{-1}$) \\ 
\hline
CXO1 & 28/11/00 & 51876.5 & Chandra/ACIS-S & 708  & 6.6 & $0.182\pm0.005$ \\
CXO2 & 19/02/01 & 51959.5 & Chandra/ACIS-S & 709  & 7.8 & $0.097\pm0.004$ \\
CXO3 & 23/03/01 & 51991.9 & Chandra/ACIS-S & 710  & 7.4 & $0.129\pm0.004$ \\
CXO4 & 20/04/01 & 52019.6 & Chandra/ACIS-S & 711  & 9.2 & $0.124\pm0.004$ \\
CXO5 & 04/05/02 & 52399.0 & Chandra/ACIS-S & 3484 & 6.5 & $0.166\pm0.005$ \\
CXO6 & 20/05/02 & 52414.4 & Chandra/ACIS-S & 3485 & 7.0 & $0.187\pm0.005$ \\
CXO7 & 11/06/02 & 52436.2 & Chandra/ACIS-S & 3486 & 6.5 & $0.349\pm0.007$ \\
CXO8 & 05/07/02 & 52460.7 & Chandra/ACIS-S & 3487 & 5.9 & $0.100\pm0.004$ \\
CXO9 & 22/07/02 & 52477.9 & Chandra/ACIS-S & 3488 & 6.5 & $0.091\pm0.004$ \\
CXO10 & 18/08/02 & 52504.4 & Chandra/ACIS-S & 3489 & 7.1 & $0.083\pm0.003$ \\
CXO11 & 03/09/02 & 52520.7 & Chandra/ACIS-S & 3490 & 6.9 & $0.107\pm0.004$ \\
XMM1 & 15/10/02 & 52562.1 & XMM-Newton/MOS & 0112440301 & 7.1  & $0.051\pm0.003$ (MOS1) \\
     &    &         &                &            &      & $0.055\pm0.003$ (MOS2) \\
XMM2 & 17/10/02 & 52564.2 & XMM-Newton/MOS & 0112440401 & 13.4 & $0.047\pm0.002$ (MOS1) \\
     &    &         &                &            &      & $0.047\pm0.002$ (MOS2) \\
XMM3 & 27/10/02 & 52574.1 & XMM-Newton/MOS & 0112440101 & 2.7  & $0.057\pm0.005$ (MOS1) \\
     &    &         &                &            &      & $0.057\pm0.005$ (MOS2) \\
\hline
\end{tabular}
\end{table*}

\subsection{Chandra data reduction}

We reduced the {\it Chandra} data using the CIAO software (v4.2) and the most
recent calibration database (CALDB v4.3.1).  All observations were performed
using the ACIS-S detector, with a 1/8 sub-array to give a frame time of
0.44s and the source placed off-axis where the PSF is significantly larger. Both these choices mitigate pile-up
\citep[see][]{rutledge02}, even for the brightest observation here. In all cases, we followed the standard data reduction
threads\footnote{see http://cxc.harvard.edu/ciao/threads/}, using the
\verb|chandra_repro| script to reprocess the data
with the latest calibration.  The source spectrum was extracted using a circular
region of radius 10 pixels, and the background spectrum was extracted from a
source-free annulus with inner radius 20 pixels and outer radius 60 pixels.  The
response matrix (rmf) and ancillary response file (arf) were generated with the
\verb|mkacisrmf| and \verb|mkarf| scripts.

\subsection{XMM-Newton data reduction}

All three {\it XMM-Newton} observations analysed here were performed with the PN
detector operated in timing mode.  As the source was in quiescence during these
observations, the timing mode data is not of a high quality and the source is only detectable when a restricted energy range is used.  We therefore analyse only the data for the two MOS detectors, which were both operated in full frame mode with the thin filter for each observation. These MOS data were
reduced using the XMMSAS software (v10.0.0), with the latest `current
calibration files', producing calibrated event lists from the Observation Data
Files using \verb|emproc|.  

There was no significant background flaring in the first (0112440301) or second
(0112440401) {\it XMM-Newton} observations analysed here.  However, during the
third observation (0112440101), there was a bright background flare
at the end of the observation.  We searched for background flares by filtering
the lightcurve from the whole detector for single events (pattern zero) with
energies above 10 keV.  We excluded times when the count rate from this
filtered lightcurve was greater than 2 counts~s$^{-1}$. This reduced the
exposure by approximately 0.5 ks.  
 
Using the \verb|evselect| tool, the source spectra were extracted from a
circular region of radius 32\arcsec, where as the background spectra were
extracted from an annulus with inner radius 60\arcsec and outer radius
250\arcsec.  We selected for events with patterns 0 -- 12 only.  The rmf and arf
were generated with the \verb|rmfgen| and \verb|arfgen| tools.

\section{Spectral Analysis}\label{sec:spec}

The spectra were modelled using the XSPEC (v12) spectral fitting
package \citep{arnaud96}.  All spectra were grouped to a minimum of 20 counts per bin in the 0.5 -- 10 keV energy range, and the models were only fit over this energy range. All uncertainties are quoted at the 1$\sigma$ confidence level.

An initial look at the raw count rates from the eleven {\it Chandra}
observations (see Table~\ref{tab:obs}) demonstrates there is variability between
the observations. For
instance, CXO7 has a count rate about 4 times higher than CXO10. 
Initially we fit all the spectra individually with an absorbed neutron star
atmosphere model, using the \verb|phabs| model for
photoelectric absorption and the \verb|nsatmos| model \citep{heinke06} for the
neutron star atmosphere. From these fits, it is clear that half of the
spectra are fit well ($\chi^2_\nu < 1.2$) by just an absorbed neutron star
atmosphere model, but that the other half require an additional power-law
component. The observations that require a power-law
component ($\chi^2_\nu > 1.5$ without it) are CXO3, CXO4, CXO5, CXO6, CXO7,
CXO11 and XMM2.  Note that in most of those fits the power-law index is poorly
constrained, and all spectra are consistent with having the same index at the
1$\sigma$ level.  In these fits where the power-law is required, the
index ranges from approximately -1 to 2, but the uncertainties are such that
they are all consistent with $\Gamma = 1$.

In order to determine which spectral parameters are variable between the
observations, we fit all the spectra jointly.  Here, we fit an absorbed
neutron star atmosphere plus power-law model.  We include the power-law
component for all observations, allowing the normalisation to be a free
parameter. Given that in the individual fits the power-law index was
consistent with being constant, we tied the power-law index between all
observations. From the individual fits, we also find that the column
density is consistent between all the observations, and therefore we tie this
parameter. We also assume a canonical neutron star mass and radius (1.4 \Msun,
10 km), and that the entire neutron star surface
is emitting (these parameters were all fixed).  We assume a distance of 5 kpc
throughout, as also adopted in previous work \citep[e.g.][]{rutledge02}.  For a
discussion of the distance to Aql~X-1 see \citet{rutledgeetal01b} who find the
distance is between 4 and 6.5 kpc.

\begin{figure}
\centering
\includegraphics[width=8.4cm]{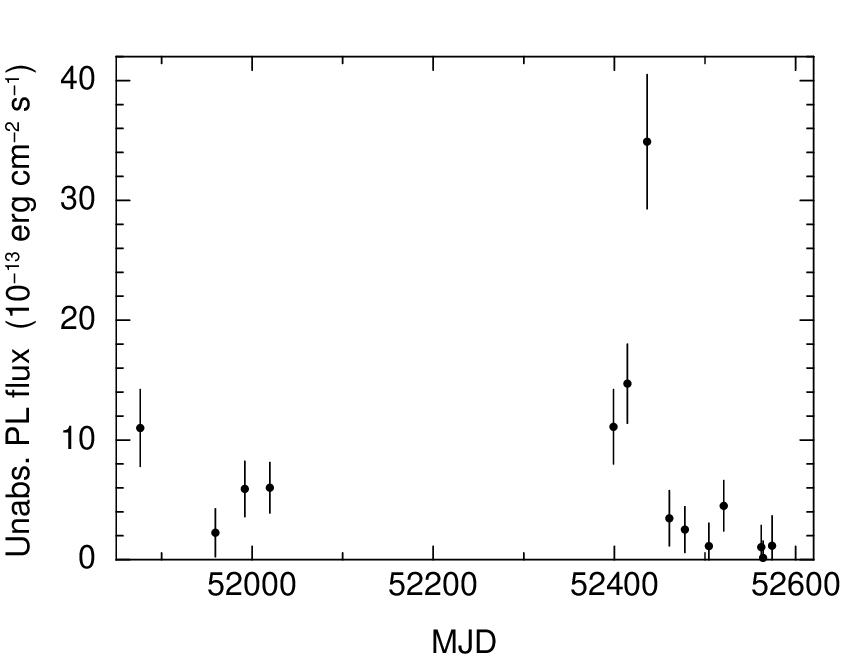}
\caption{Variability of the unabsorbed 0.5 -- 10 keV power-law flux when this is
the only parameter allowed to vary between observations (the effective
temperature and power-law index are tied between the observations).}
\label{fig:qlc_pl}
\end{figure}

In the first instance, we tie the neutron star atmosphere temperature between
all observations (the spectral parameters are given in
Table~\ref{tab:specfit} and all uncertainties quoted there, and throughout
the paper, are at the 1$\sigma$ confidence level). This allowed us to
investigate whether variability in the power-law normalisation alone can fit the
data.  This provides an adequate fit to the data ($\chi^2_\nu = 1.10, \nu = 533$, null hypothesis probability = 0.057),
with an effective temperature (for an observer at infinity) of $kT_{\rm
eff}^{\infty} = 108.9\pm0.8$ eV, power-law index, $\Gamma = 2.73\pm0.06$ and
power-law normalisation ranging from a non-detection in XMM2 to approximately
$10^{-3}$ photons keV$^{-1}$ cm$^{-2}$ s$^{-1}$ at 1 keV in CXO7. The
variability of the unabsorbed 0.5 -- 10 keV power-law flux is shown in
Figure~\ref{fig:qlc_pl} for this model.  Note that the column density we
determine is consistent with values determined from previous work, as
well as optical photometry and 21 cm emission \citep[see][ for a discussion of
the reddening towards Aql X-1]{rutledge02}.

\begin{table*}
\centering
\caption{Spectral fitting parameters.  The spectra were fit jointly, with the column density and power-law index tied between all observations in all cases.  The neutron star mass and radius were fixed at canonical values (1.4~M$_\odot$, 10~km). The power-law normalisation is in units of photons keV$^{-1}$ cm$^{-2}$ s$^{-1}$ and is defined at 1 keV.  The luminosity, $L$, is evaluated over the 0.5 -- 10 keV range and is for $D = 5$ kpc. The uncertainty in the luminosity is generally dominated by the uncertainty in $D$.}
\label{tab:specfit}
{\scriptsize
\begin{tabular}{ccccccccc}
\hline
Obs. & $N_{\rm H}$ & $kT_{\rm eff}^{\infty}$ & $\Gamma$ & Power law norm. & 0.5 -- 10 keV flux & $L$ & Thermal & $\chi_\nu^2$ (dof) \\
 & ($10^{21}$ cm$^{-2}$) & (eV) & & ($10^{-5}$) & ($10^{-13}$ erg cm$^{-2}$ s$^{-1}$) & ($10^{33}$ erg s$^{-1}$) & fraction &\\ 
\hline
\multicolumn{9}{c}{Power-law normalisation variable, temperature tied} \\
\hline
CXO1 & $4.3\pm0.1$ & $108.9\pm0.8$ & $2.73\pm0.06$ & $34.2\pm3.2$ & $8.0\pm0.6$ & $5.6\pm2.7$ & $0.41\pm0.03$ & 1.10 (533)\\
CXO2 & & & & $7.0\pm2.0$ & $3.8\pm0.4$ & $2.9\pm1.4$ & $0.77\pm0.09$ & \\
CXO3 & & & & $18.3\pm2.3$ & $5.5\pm0.5$ & $4.0\pm2.0$ & $0.56\pm0.05$ & \\
CXO4 &  & & & $18.6\pm2.1$ & $5.6\pm0.5$ & $4.1\pm2.0$ & $0.56\pm0.05$ & \\
CXO5 &  & & & $34.5\pm3.1$ & $8.0\pm0.7$ & $5.6\pm2.7$ & $0.40\pm0.04$ & \\
CXO6 &  & & & $45.3\pm3.3$ & $9.7\pm0.6$ & $6.7\pm3.2$ & $0.34\pm0.02$ & \\
CXO7 &  & & & $107.8\pm5.6$ & $19.4\pm0.9$ & $12.7\pm6.1$ & $0.18\pm0.01$ &\\
CXO8 &  & & & $10.7\pm2.3$ & $4.4\pm0.6$ & $3.3\pm1.6$ & $0.68\pm0.10$ &\\
CXO9 &  & & & $7.8\pm1.9$ & $3.9\pm0.5$ & $3.0\pm1.5$ & $0.75\pm0.10$ &\\
CXO10 &  & & & $3.5\pm1.9$ & $3.3\pm0.5$ & $2.6\pm1.3$ & $0.87\pm0.11$ &\\
CXO11 &  & & & $13.9\pm2.1$ & $4.9\pm0.6$ & $3.6\pm1.8$ & $0.63\pm0.08$ &\\
XMM1 &  & & & $3.3\pm1.8$ & $3.2\pm0.4$ & $2.6\pm1.3$ & $0.87\pm0.11$ &\\
XMM2 &  & & & $0.5^{+1.4}_{-0.5}$ & $2.8\pm0.4$ & $2.3\pm1.2$ & $0.97_{-0.14}^{+0.03}$ & \\
XMM3 &  & & & $3.6\pm2.5$ & $3.3\pm0.5$ & $2.6\pm1.3$ & $0.86\pm0.13$ &\\
\hline
\multicolumn{9}{c}{Temperature and power-law normalisation variable} \\
\hline
CXO1 & $3.8\pm0.1$ & $126.4\pm0.9$ & $0.80\pm0.25$ & $0.9\pm0.3$ & $8.0\pm0.8$ & $4.9\pm2.4$ & $0.89\pm0.09$ & 0.88 (520)\\
CXO2 & & $109.3\pm1.3$ & & $1.2\pm0.6$ & $5.2\pm2.1$ & $3.1\pm1.9$ & $0.76_{-0.31}^{+0.24}$ & \\
CXO3 & & $116.1\pm1.0$ & & $1.4\pm0.4$ & $6.8\pm1.2$ & $3.9\pm2.0$ & $0.78\pm0.14$ &\\
CXO4 & & $114.6\pm1.0$ & & $1.9\pm0.7$ & $7.5\pm1.2$ & $4.0\pm2.0$ & $0.71\pm0.12$ &\\
CXO5 & & $124.2\pm1.1$ & & $1.7\pm0.6$ & $9.0\pm1.3$ & $5.1\pm2.6$ & $0.80\pm0.12$ &\\
CXO6 & & $125.2\pm1.1$ & & $3.5\pm1.3$ & $12.6\pm2.0$ & $6.4\pm3.2$ & $0.67\pm0.11$ &\\
CXO7 & & $142.4\pm1.3$ & & $7.2\pm2.6$ & $25.1\pm4.1$ & $11.8\pm6.0$ & $0.63\pm0.11$ &\\
CXO8 & & $112.2\pm1.4$ & & $1.0\pm0.6$ & $5.3\pm2.1$ & $3.2\pm2.0$ & $0.81_{-0.32}^{+0.19}$ &\\
CXO9 & & $110.1\pm1.2$ & & $0.9\pm0.3$ & $4.9\pm1.0$ & $3.0\pm1.6$ & $0.80\pm0.17$ & \\
CXO10 & & $107.1\pm1.4$ & & $1.2\pm0.5$ & $5.0\pm1.6$ & $2.8\pm1.6$ & $0.74\pm0.24$ &\\
CXO11 & & $113.4\pm1.1$ & & $1.2\pm0.4$ & $6.0\pm0.9$ & $3.5\pm1.8$ & $0.79\pm0.12$ &\\
XMM1 &  & $107.9\pm1.0$ & & $0.6\pm0.2$ & $3.9\pm0.8$ & $2.5\pm1.3$ & $0.86_{-0.18}^{+0.14}$ &\\
XMM2 &  & $106.2\pm0.9$ & & $0.5\pm0.2$ & $3.5\pm0.7$ & $2.3\pm1.2$ & $0.88_{-0.18}^{+0.12}$ &\\
XMM3 &  & $109.9\pm1.1$ & & $0.0^{+0.8}$ & $3.1\pm1.7$ & $2.4\pm1.7$ & $1.00_{-0.55}$ & \\
\hline
\end{tabular}}
\end{table*}

\begin{figure}
\centering
\includegraphics[angle=270,width=8.4cm]{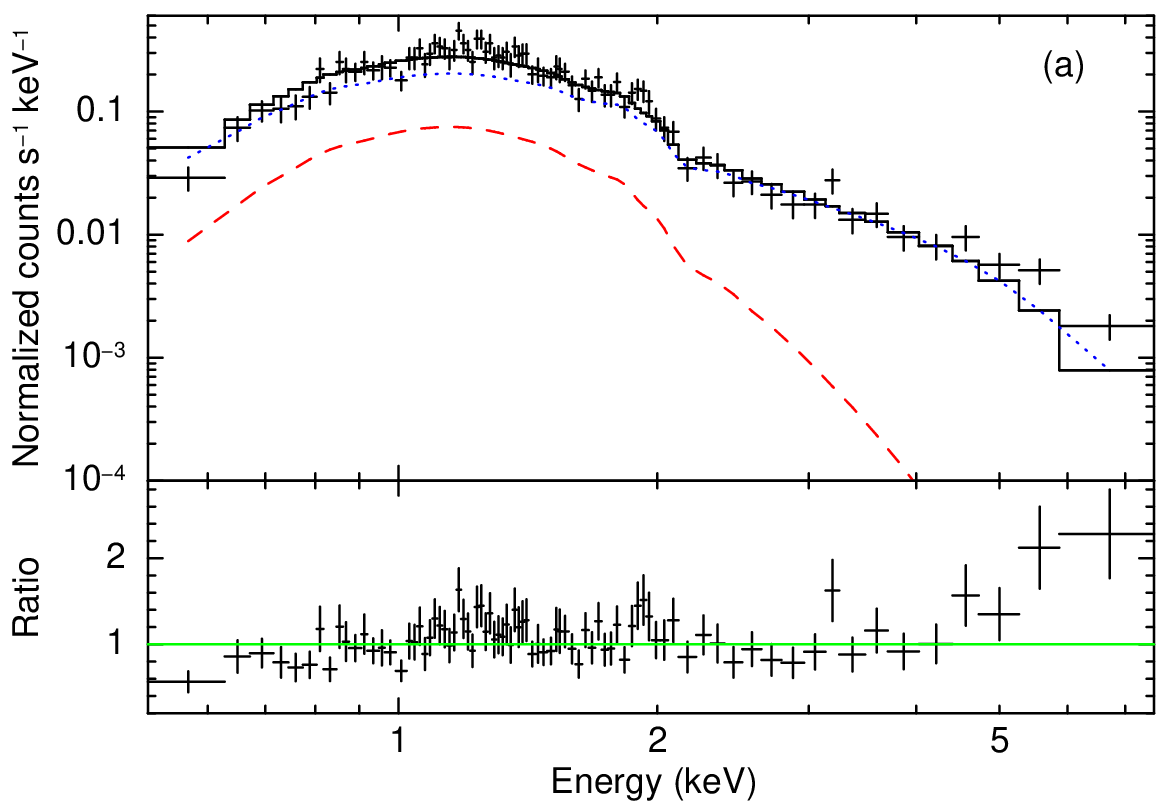}
\includegraphics[angle=270,width=8.4cm]{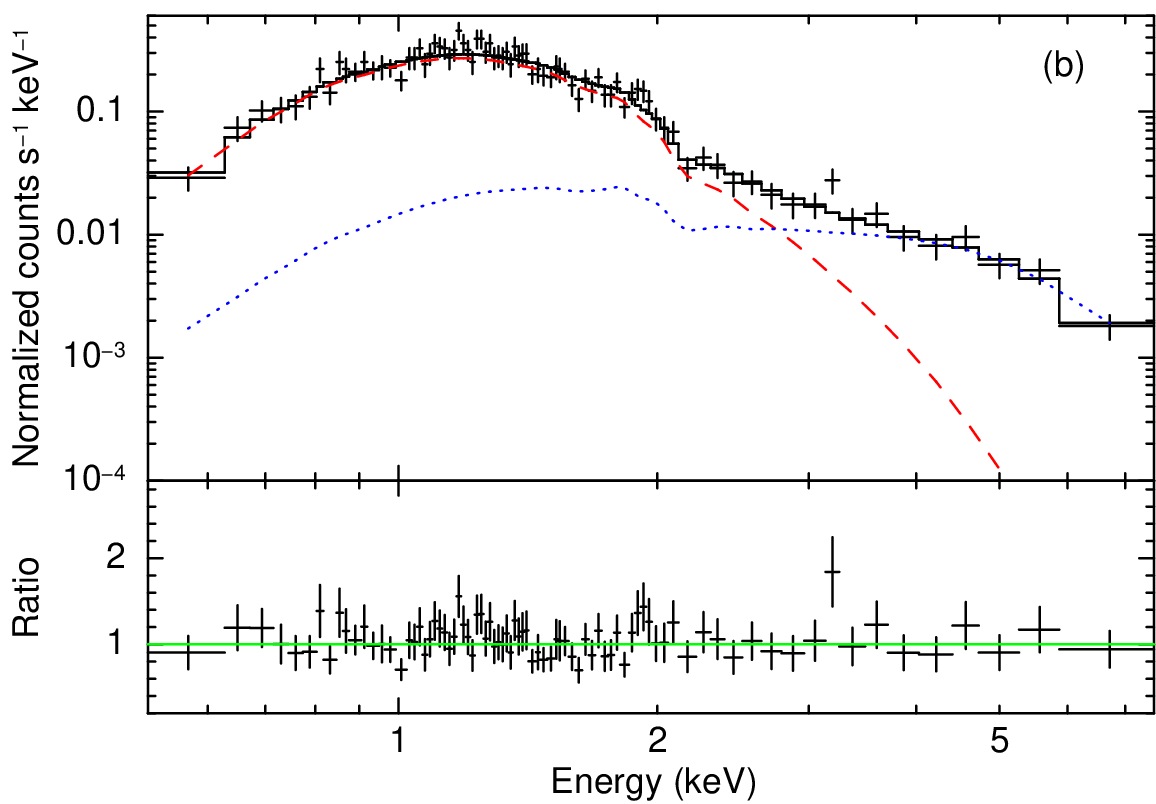}
\caption{The brightest {\it Chandra} spectrum of Aql X-1, CXO7. Panel (a)
displays the model from jointly fitting all spectra with only the power-law
normalisation variable.  Panel (b) shows the model from jointly fitting all
spectra with both the effective temperature of the neutron star atmosphere and
the power-law normalisation variable.  In both cases the dashed red line marks
the neutron star atmosphere component, and the dotted blue line marks the
power-law component.}
\label{fig:spec}
\end{figure}

\begin{figure}
\centering
\includegraphics[width=8.4cm]{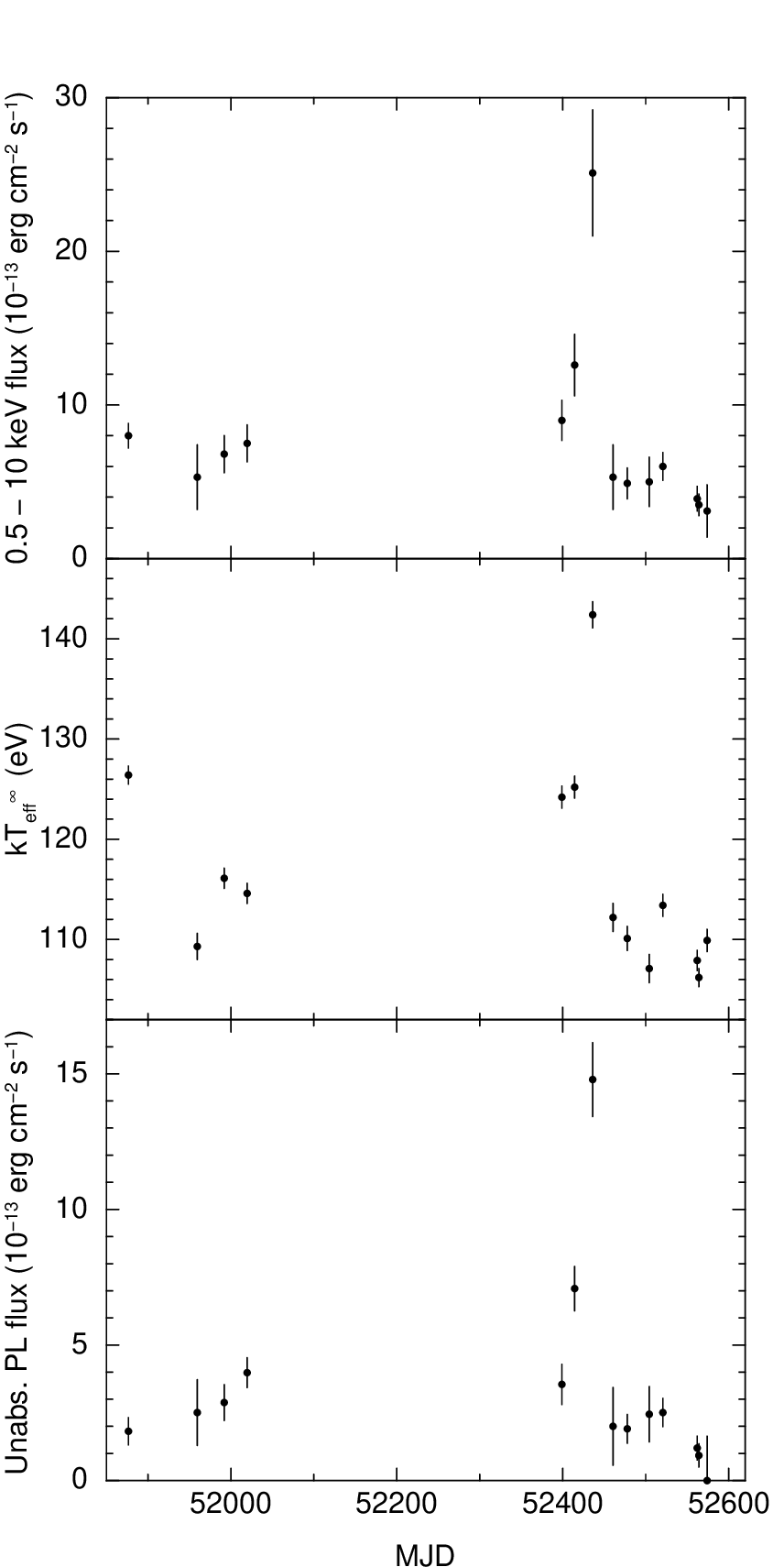}
\caption{Quiescent lightcurve of Aql~X-1 for the model where temperature and
power-law normalisation are allowed to vary.  The top panel shows the 0.5 -- 10
keV flux, the middle panel the effective temperature (for an observer at
infinity), and the bottom panel the unabsorbed 0.5 -- 10 keV power-law
flux.}
\label{fig:qlc}
\end{figure}

While the model with only the power-law normalisation variable between
observations fits the data well, examining the residuals of the brightest
observation (CXO7) it seems that the steep power-law is attempting to
mimic a soft thermal component, and the residuals show an upturn above 4 keV
(see Figure~\ref{fig:spec}). Note that while this is not statistically
significant, it would appear to suggest the spectrum should be dominated by a
thermal component at soft energies rather than the power-law.  We therefore also
fit the data allowing both the power-law normalisation and the temperature of
the neutron star atmosphere to vary.  The results of those fits are also given
in Table~\ref{tab:specfit}.  A good fit, with a lower $\chi^2$, is achieved
($\chi^2_\nu = 0.88, \nu = 520$).  In this case a much flatter power-law index
is found, $\Gamma = 0.80\pm0.25$.  A power-law consistent with a
slope of 1 -- 2 is quite typical for quiescent neutron stars.
With this model we find significant variability in the effective
temperature of the neutron star atmosphere component, ranging from approximately
$kT_{\rm eff}^{\infty} = 106 - 142$ eV (with typical 1$\sigma$ uncertainties of
1 eV), in addition to variability in the power-law normalisation.  In
Figure~\ref{fig:qlc} we show the 0.5 -- 10 keV flux, effective temperature and
unabsorbed 0.5 -- 10 keV power-law flux from this model.  For comparison, we show
the spectral fit to CXO7 for both models in Figure~\ref{fig:spec}, note that the
residuals are much flatter at both high and low energy in the fit where both the
temperature and the power-law normalisation are allowed to vary.

We also investigated a fit with both the power-law index and normalisation tied
between all observations.  It provides a significantly worse fit ($\chi^2_\nu = 1.16, \nu = 533$, null hypothesis probability = $6.6\times10^{-3}$), and cannot match the spectrum of the brightest observations above 3 keV,  underpredicting the flux at 5 keV by a factor of $\sim4$.  As discussed in the Introduction, \citet{campanastella03} described the first four {\it Chandra}
observations by correlated changes in the power-law index and column density. 
Note that when fitting the spectra individually, the $N_{\rm H}$ values are all
consistent with remaining unchanged, and as mentioned above, so are the
power-law indices.  Moreover, given that a good fit can be achieved without the
power-law index or $N_{\rm H}$ being variable, we do not test this model
further.  

The three {\it XMM-Newton} observations have some of the lowest 0.5 -- 10 keV fluxes of all the observations.  While this change may be real, we caution that part of the difference in flux with the closest {\it Chandra} observations could
potentially be due to cross-calibration differences between the two missions. 
Differences of 10\% between {\it Chandra} and {\it XMM-Newton} soft X-ray fluxes
have been seen in cross-calibration studies \citep[e.g.][]{tsujimoto11}.

We also investigated the effect of having fit all the data simultaneously with
multiple parameters tied between the spectra.  With such a method, the spectrum
with the highest signal-to-noise ratio can potentially dominate the fit, and
skew the parameters.  However, we find that this is not the case here. Removing
the best spectrum, CXO7, from the fit with both temperature and power-law
normalisation variable, we find that both tied parameters (the power-law index
and $N_{\rm H}$) remain consistent with their former values, with
$\Gamma=0.79\pm0.35$ and $N_{\rm H} = (3.8\pm0.1 )\times10^{21}$ cm$^{-2}$.

As an aside, we note that we searched for any obvious variability or flaring
within the individual lightcurves of each observation.  There is no dramatic
variability present \citep[though see][ for power density spectra of the
first four {\it Chandra} observations]{rutledge02}, and we do not investigate
this further here.

\section{Discussion}\label{sec:disc}

We have analysed 14 observations of Aql~X-1 in quiescence, with the
observations spanning a period of approximately 2 years.  Variability in the
first four observations has been studied previously
\citep{rutledge02,campanastella03}.  Here, we find variability between the
different observations, with a particularly striking flare showing a change in
0.5 -- 10 keV flux by a factor of $\sim$5 between the peak of the flare (CXO7)
and the next observation (CXO8). If the three observations around MJD 52400 are
from the same event, then the flare lasted approximately 60 days, with
observations CXO5 and CXO6 seemingly catching the rise of the flare.  However,
with the sparse sampling we cannot be sure that this was a single flare event
or not.  The ratio of the maximum to minimum observed 0.5 -- 10 keV flux is 7 --
8 (depending on the spectral model assumed).

We fitted the spectra with an absorbed neutron star atmosphere plus power-law
model.  In the first instance, the only parameter allowed to vary between the
observations was the power-law normalisation, i.e. we assume that the
temperature of the neutron star surface remains the same at all epochs.  Such a
model fits the data adequately ($\chi^2_\nu = 1.10$), and requires a power-law with a steep slope ($\Gamma = 2.7$). Alternatively, we also fit the data
allowing both the effective temperature and power-law normalisation vary, which,
of course, also fits well ($\chi^2_\nu = 0.88$).  In this case, the power-law
index is much flatter ($\Gamma = 0.8$), and the thermal component dominates the
spectrum at all epochs (with thermal fractions typically 0.7 -- 0.9).  Given
that both models provide a good fit to the data, we are unable to conclusively
determine which spectral component is driving the variability.

Comparing the times of the quiescent observations with the {\it RXTE}/ASM
lightcurve (Fig~\ref{fig:asmlc}), the last significant (3$\sigma$) detection at
the end of the outburst (before the flare) with the {\it RXTE}/ASM was on MJD
52355. A pointed {\it RXTE}/PCA observation detects Aql~X-1 on MJD 52366 with a
2 -- 10 keV flux of $8.5^{+0.8}_{-3.3}\times10^{-12}$ erg s$^{-1}$ cm$^{-2}$,
however, a subsequent (and final) pointed observation two days later does not
provide a clear detection, and gives an upper limit to the 2 -- 10 keV flux of
approximately $5.5\times10^{-12}$ erg s$^{-1}$ cm$^{-2}$. Therefore, the first
of the three bright observations (CXO5, 6 and 7) occurs about 33 days after the
end of a full outburst (taking the last detection with the PCA as the
approximate end of the outburst).

\begin{figure}
\centering
\includegraphics[width=8.0cm]{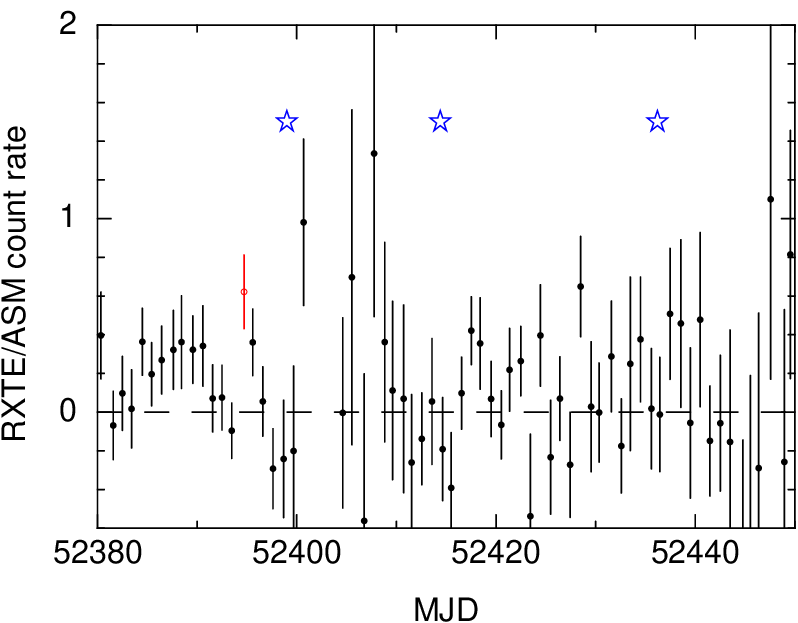}
\caption{The RXTE/ASM 1-day averaged lightcurve of Aql X-1 around the times of
the three brightest quiescent observations.  There is no apparent flaring
detected in the ASM lightcurve. The markers are as in Figure~\ref{fig:asmlc}.}
\label{fig:zoom}
\end{figure}

Figure~\ref{fig:zoom} shows the {\it RXTE}/ASM 1-day averaged lightcurve around
the times of the three brightest quiescent observations.  Clearly, there is no
apparent flaring detected by the ASM. There is only one data point at a
significance greater than 3$\sigma$, and that occurs approximately 5 days before
CXO5 (the observation on MJD 52399).  Given the surrounding non-detections, this
point is likely a statistical fluctuation and not a real detection. Even so,
Aql~X-1 was not accreting at close to outburst levels near the quiescent flare
we observed.

One source of variability in quiescence is thermal relaxation of the neutron star crust after the end of an outburst \citep{rutledge_ks1731_02}, which has now been observed in four sources \citep[e.g.][]{cackett06,fridriksson10,degenaar10}.  As CXO5, 6 and 7 occur close to the end of an outburst, it may be possible
that thermal relaxation of the crust could lead to
variability in the thermal component.  Thus far, only an overall luminosity
decrease over time has been observed. However, depending on the microphysics of
the crust and core, an increase can be achieved as heat from pycnonuclear
reactions diffuses to the top of the crust
\citep{ushomirskyrutledge01,rutledge_ks1731_02}. The largest
fractional variability would come from sources with long recurrence times
between outbursts and from neutron stars with enhanced levels of core neutrino
emission. But, as noted by \citet{ushomirskyrutledge01}, the quiescent
luminosity of Aql X-1 is inconsistent with rapid cooling \citep{BBR98} and it
has a short recurrence time ($\sim1$ yr), thus the observed quiescent
variability in this source cannot be due to thermal relaxation of the crust.
Furthermore, the observed flare is not just due to the thermal component - the
thermal fraction is lowest during the flare and there is a significant and
clearly variable power-law component. Note also that CXO1 also occurs directly
after the end of an outburst, and is hotter than the 3 subsequent observations.
 While it is tempting to speculate that the neutron star was hotter in CXO1
than CXO 2 -- 4 because of thermal relaxation of the crust, the arguments of
\citet{ushomirskyrutledge01} (discussed above), would seem to rule that out.

While thermal relaxation of the crust does not appear to explain the flare, it may still be possible that the flare is associated with the previous outburst.  A double outburst was recently seen from the accretion-powered millisecond pulsar IGR~J00291+5934 \citep{patruno10,hartman11,papitto11}, where two approximately 10 day long outbursts were separated by only 30 days in quiescence.  Discussing these two outbursts, \citet{hartman11} suggest that they are connected -- the first outburst is stopped by the propeller effect before the disc is completely depleted, leaving material in the disc for the second outburst.  The first brighter observation in Aql X-1 is also approximately 30 days after the end of an outburst, however, unlike IGR~J00291+5934 the increase in Aql~X-1 does not get close to outburst levels (Aql~X-1 is not detected by the RXTE/ASM near the brightest observations), thus it is not clear whether a similar mechanism applies here.

In addition to Aql~X-1, flaring (by which we mean short lasting, sharp rises in
flux) during quiescence has recently been observed in
the neutron star low-mass X-ray binaries XTE~J1701$-$462
\citep{fridriksson10} and Cen~X-4 \citep{cackett10}.  In XTE~J1701$-$462, an
increase in X-ray flux by a factor of almost 3 (to a peak observed 0.5 -- 10 keV
luminosity of $2.65\times10^{34}$ erg s$^{-1}$) was observed approximately
230 days after the end of an outburst on top of an overall decreasing X-ray flux
presumably associated with thermal relaxation of the crust
\citep{fridriksson10}.  The increased flux lasted a maximum of approximately
120 days (the time between the quiescent observations on either side of the
increased flux). Since then, further X-ray monitoring of this source during
quiescence showed another X-ray flare about 1075 days into quiescence.  This
time, the flare was significantly brighter, with a peak observed increase of a
factor of $\sim$20 in flux (reaching a maximum luminosity of
$\sim1\times10^{35}$ erg s$^{-1}$) and lasting 10 -- 20 days in total
\citep{fridriksson11}. Interestingly, like Aql X-1, in XTE~J1701$-$462 the
thermal component does increase during the flares, but the non-thermal component
increases faster leading to a thermal fraction that is lower during the flares
than at other times in quiescence \citep{fridriksson11}.

In Cen~X-4, quiescent observations spanning 15 years showed substantial
variability between epochs \citep{cackett10}. Additionally, examination of the
lightcurves from individual observations revealed a short flare in one of the
{\it XMM-Newton} observations. This flare lasted $\sim$2 ks and showed an
increase of approximately a factor of 4 in count rate at the peak. The natural
explanation for this flaring activity in both these sources appears to be
increased levels of ongoing accretion during quiescence, which presumably is
occurring in Aql X-1 too.  The exact mechanism for this accretion is unclear (as
discussed in the Introduction, the origin of the power-law component is
uncertain), but the clear variability in the thermal component in Cen X-4, and
the apparent connection between the power-law and thermal components (the
thermal fraction remained roughly constant) may suggest that accreting material
reaches the neutron star surface.

Another class of objects that are particularly interesting are the very faint
X-ray transients (VFXTs) whose peak 2 -- 10 keV luminosity during outburst only
reaches $10^{34 - 36}$ erg s$^{-1}$.  With such low peak luminosities, these
sources are missed by all-sky monitors.  However, frequent monitoring of the
Galactic Centre region by multiple X-ray missions (e.g. {\it Chandra},
{\it XMM-Newton} and {\it Swift}), has revealed a number of these sources
\citep[see, e.g.][]{muno05,sakano05,wijnands06,degenaar09b,degenaar10b}.  The
frequent monitoring has revealed flare-like behaviour in-between normal
quiescent and outburst levels in four sources \citep{degenaar09b,degenaar10b}. 
The flares last typically 1 -- 2 weeks, and the peak brightness can vary
substantially, though one source (XMM~J174457$-$2850.3) has shown a flare with a
peak luminosity as low as seen here in Aql X-1.  Several (though not all) of
these VFXTs are confirmed as neutron star sources as they have shown type-I
X-ray bursts.

Variability in quiescence is not limited to just neutron stars. 
Several of the brighter quiescent stellar-mass black holes have also been
observed to vary in several sources \citep[V404~Cyg, 4U~1630$-$47, GX~339$-$4,
A0620$-$00, and
GRO~J1655$-$40:][]{wagner94,parmar97,kong00,kong02,hynes04,bradley07,
millerjones08}. The best studied of these is V404~Cyg.  This object has been
observed to vary by as much as a factor of 20 in X-ray count rate during one
short flare \citep{hynes04}, and has been variable during the observation on
every occasion when observed by {\it Chandra} or {\it XMM-Newton}
\citep{kong02, hynes04, bradley07}.  Also interesting is that the X-ray
variability is correlated with optical variability, with the H$\alpha$ line
variability seemingly powered by X-ray irradiation \citep{hynes04}.  Moreover,
sensitive radio observations of V404~Cyg in quiescence detect the source and
even find a short flare, with a rise time of $\sim$30 minutes
\citep{millerjones08}.  The radio emission appears to be non-thermal and
therefore may originate in a compact jet.  Of course, in these objects there is
no stellar surface, thus variability here cannot involve any stellar surface or
stellar magnetic field.  While it has been suggested that coronal emission from
the rapidly rotating secondary star could be the source of the quiescent X-ray
emission in black hole sources \citep{bildstenrutledge00}, the X-ray luminosity
of several sources exceeds the maximum predicted by the coronal model
\citep[e.g.][]{kong02}.

Quiescent variability and flaring has also been seen in nearby quiescent
supermassive black holes. For instance, the supermassive black hole at the
centre of the Galaxy, Sgr A*, has shown significant flaring behaviour at X-ray,
IR and radio wavelengths
\citep[e.g.][]{baganoff01,genzel03,zhao03,porquet03,porquet08}.  The
brightest X-ray flare from Sgr A* showed an increase by a factor of 160 in 2-10
keV luminosity with a duration shorter than one hour \citep{porquet03}. More
recently, the supermassive black hole at the centre of Andromeda,
M31*, has also shown X-ray variability and flaring activity
\citep{garcia10,ligarciaetal10}.  Thus, while any mechanism for such flares is
not clear \citep[e.g.][ and references therein]{yuan04,maitra09}, it may suggest
that flaring is a standard property of accretion at low rates, though the
stellar surface and magnetic field of neutron stars may lead to a different
inner accretion geometry. 

The standard model used to explain the transient behaviour and outburst cycles in X-ray binaries is the disc instability model \citep[e.g.][]{kingritter98,hameury98,menou00,dubus01,lasota01}.  During quiescence, as long as the mass accretion rate is smaller than some critical value everywhere in the disc, accretion can continue \citep*[see for example the discussion in][]{kuulkers09}.  However, the disc instability model does not make any strong predictions for variability during quiescence, and \citet{dubus01} discuss how flaring after the end of an outburst such as that observed in GRO J0422+32 \citep{callanan95} are not understood.  Similarly, it appears that the variability seen in Aql X-1 and XTE J1701-462 can not readily be explained in the disc instability framework.

Finally, the short 2 ks flare seen in the X-ray lightcurve of Cen~X-4 may have a different origin (not related to the disc instability model).  Such short 30 -- 60 min flares are also seen in the optical lightcurve of Cen X-4 \citep{shahbaz10}.  \citet{shahbaz10} suggest that the most likely model for those flares is blackbody radiation from an optically thin layer of recombining hydrogen (similar to the model for solar flares), and must occupy a small (0.3 per cent) area of the disc.  Further monitoring of quiescent X-ray binaries is required to understand the prevalence of such variability and flares.

\section*{Acknowledgements} EMC thanks Andy Fabian, Jim Pringle
and Ramesh Narayan for interesting and insightful conversations about accretion
at low rates. RW acknowledges support from a European Research Council Starting
Grant.

\bibliographystyle{mn2e}
\bibliography{qNS}

\begin{thebibliography}{}

\bibitem[\protect\citeauthoryear{{Arnaud}}{{Arnaud}}{1996}]{arnaud96}
{Arnaud} K.~A.,  1996, in ASP Conf. Ser. 101: Astronomical Data Analysis
  Software and Systems V {XSPEC: The First Ten Years}.
p.~17

\bibitem[\protect\citeauthoryear{{Baganoff}, {Bautz}, {Brandt}, {Chartas},
  {Feigelson}, {Garmire}, {Maeda}, {Morris}, {Ricker}, {Townsley} \&
  {Walter}}{{Baganoff} et~al.}{2001}]{baganoff01}
{Baganoff} F.~K.,  {Bautz} M.~W.,  {Brandt} W.~N.,  {Chartas} G.,  {Feigelson}
  E.~D.,  {Garmire} G.~P.,  {Maeda} Y.,  {Morris} M.,  {Ricker} G.~R.,
  {Townsley} L.~K.,    {Walter} F.,  2001, \nat, 413, 45

\bibitem[\protect\citeauthoryear{{Bildsten} \& {Rutledge}}{{Bildsten} \&
  {Rutledge}}{2000}]{bildstenrutledge00}
{Bildsten} L.,  {Rutledge} R.~E.,  2000, \apj, 541, 908

\bibitem[\protect\citeauthoryear{{Bradley}, {Hynes}, {Kong}, {Haswell},
  {Casares} \& {Gallo}}{{Bradley} et~al.}{2007}]{bradley07}
{Bradley} C.~K.,  {Hynes} R.~I.,  {Kong} A.~K.~H.,  {Haswell} C.~A.,  {Casares}
  J.,    {Gallo} E.,  2007, \apj, 667, 427

\bibitem[\protect\citeauthoryear{{Brown}, {Bildsten} \& {Rutledge}}{{Brown}
  et~al.}{1998}]{BBR98}
{Brown} E.~F.,  {Bildsten} L.,    {Rutledge} R.~E.,  1998, \apjl, 504, L95

\bibitem[\protect\citeauthoryear{{Cackett}, {Brown}, {Miller} \&
  {Wijnands}}{{Cackett} et~al.}{2010}]{cackett10}
{Cackett} E.~M.,  {Brown} E.~F.,  {Miller} J.~M.,    {Wijnands} R.,  2010,
  \apj, 720, 1325

\bibitem[\protect\citeauthoryear{{Cackett}, {Wijnands}, {Heinke}, {Edmonds},
  {Lewin}, {Pooley}, {Grindlay}, {Jonker} \& {Miller}}{{Cackett}
  et~al.}{2005}]{cackett05}
{Cackett} E.~M.,  {Wijnands} R.,  {Heinke} C.~O.,  {Edmonds} P.~D.,  {Lewin}
  W.~H.~G.,  {Pooley} D.,  {Grindlay} J.~E.,  {Jonker} P.~G.,    {Miller}
  J.~M.,  2005, \apj, 620, 922

\bibitem[\protect\citeauthoryear{{Cackett}, {Wijnands}, {Linares}, {Miller},
  {Homan} \& {Lewin}}{{Cackett} et~al.}{2006}]{cackett06}
{Cackett} E.~M.,  {Wijnands} R.,  {Linares} M.,  {Miller} J.~M.,  {Homan} J.,
   {Lewin} W.~H.~G.,  2006, \mnras, 372, 479

\bibitem[\protect\citeauthoryear{{Callanan}, {Garcia}, {McClintock}, {Zhao},
  {Remillard}, {Bailyn}, {Orosz}, {Harmon} \& {Paciesas}}{{Callanan}
  et~al.}{1995}]{callanan95}
{Callanan} P.~J.,  {Garcia} M.~R.,  {McClintock} J.~E.,  {Zhao} P.,
  {Remillard} R.~A.,  {Bailyn} C.~D.,  {Orosz} J.~A.,  {Harmon} B.~A.,
  {Paciesas} W.~S.,  1995, \apj, 441, 786

\bibitem[\protect\citeauthoryear{{Campana}, {Colpi}, {Mereghetti}, {Stella} \&
  {Tavani}}{{Campana} et~al.}{1998}]{campanaetal98a}
{Campana} S.,  {Colpi} M.,  {Mereghetti} S.,  {Stella} L.,    {Tavani} M.,
  1998, \aapr, 8, 279

\bibitem[\protect\citeauthoryear{{Campana}, {Israel}, {Stella}, {Gastaldello}
  \& {Mereghetti}}{{Campana} et~al.}{2004}]{campanaetal04}
{Campana} S.,  {Israel} G.~L.,  {Stella} L.,  {Gastaldello} F.,    {Mereghetti}
  S.,  2004, \apj, 601, 474

\bibitem[\protect\citeauthoryear{{Campana}, {Mereghetti}, {Stella} \&
  {Colpi}}{{Campana} et~al.}{1997}]{campana97}
{Campana} S.,  {Mereghetti} S.,  {Stella} L.,    {Colpi} M.,  1997, \aap, 324,
  941

\bibitem[\protect\citeauthoryear{{Campana} \& {Stella}}{{Campana} \&
  {Stella}}{2003}]{campanastella03}
{Campana} S.,  {Stella} L.,  2003, \apj, 597, 474

\bibitem[\protect\citeauthoryear{{Degenaar} \& {Wijnands}}{{Degenaar} \&
  {Wijnands}}{2009}]{degenaar09b}
{Degenaar} N.,  {Wijnands} R.,  2009, \aap, 495, 547

\bibitem[\protect\citeauthoryear{{Degenaar} \& {Wijnands}}{{Degenaar} \&
  {Wijnands}}{2010}]{degenaar10b}
{Degenaar} N.,  {Wijnands} R.,  2010, \aap, 524, A69

\bibitem[\protect\citeauthoryear{{Degenaar}, {Wolff}, {Ray}, {Wood}, {Homan},
  {Lewin}, {Jonker}, {Cackett}, {Miller}, {Brown} \& {Wijnands}}{{Degenaar}
  et~al.}{2010}]{degenaar10}
{Degenaar} N.,  {Wolff} M.~T.,  {Ray} P.~S.,  {Wood} K.~S.,  {Homan} J.,
  {Lewin} W.~H.~G.,  {Jonker} P.~G.,  {Cackett} E.~M.,  {Miller} J.~M.,
  {Brown} E.~F.,    {Wijnands} R.,  2010, MNRAS, in press, arXiv:1007.0247

\bibitem[\protect\citeauthoryear{{Dubus}, {Hameury} \& {Lasota}}{{Dubus}
  et~al.}{2001}]{dubus01}
{Dubus} G.,  {Hameury} J.,    {Lasota} J.,  2001, \aap, 373, 251

\bibitem[\protect\citeauthoryear{{Fridriksson} et~al.,}{{Fridriksson}
  et~al.}{2011}]{fridriksson11}
{Fridriksson} J.~K.,  et~al., 2011, \apj, submitted, arXiv:1101.0081

\bibitem[\protect\citeauthoryear{{Fridriksson}, {Homan}, {Wijnands},
  {M{\'e}ndez}, {Altamirano}, {Cackett}, {Brown}, {Belloni}, {Degenaar} \&
  {Lewin}}{{Fridriksson} et~al.}{2010}]{fridriksson10}
{Fridriksson} J.~K.,  {Homan} J.,  {Wijnands} R.,  {M{\'e}ndez} M.,
  {Altamirano} D.,  {Cackett} E.~M.,  {Brown} E.~F.,  {Belloni} T.~M.,
  {Degenaar} N.,    {Lewin} W.~H.~G.,  2010, \apj, 714, 270

\bibitem[\protect\citeauthoryear{{Garcia}, {Hextall}, {Baganoff}, {Galache},
  {Melia}, {Murray}, {Primini}, {Sjouwerman} \& {Williams}}{{Garcia}
  et~al.}{2010}]{garcia10}
{Garcia} M.~R.,  {Hextall} R.,  {Baganoff} F.~K.,  {Galache} J.,  {Melia} F.,
  {Murray} S.~S.,  {Primini} F.~A.,  {Sjouwerman} L.~O.,    {Williams} B.,
  2010, \apj, 710, 755

\bibitem[\protect\citeauthoryear{{Genzel}, {Sch{\"o}del}, {Ott}, {Eckart},
  {Alexander}, {Lacombe}, {Rouan} \& {Aschenbach}}{{Genzel}
  et~al.}{2003}]{genzel03}
{Genzel} R.,  {Sch{\"o}del} R.,  {Ott} T.,  {Eckart} A.,  {Alexander} T.,
  {Lacombe} F.,  {Rouan} D.,    {Aschenbach} B.,  2003, \nat, 425, 934

\bibitem[\protect\citeauthoryear{{Hameury}, {Menou}, {Dubus}, {Lasota} \&
  {Hure}}{{Hameury} et~al.}{1998}]{hameury98}
{Hameury} J.,  {Menou} K.,  {Dubus} G.,  {Lasota} J.,    {Hure} J.,  1998,
  \mnras, 298, 1048

\bibitem[\protect\citeauthoryear{{Hartman}, {Galloway} \&
  {Chakrabarty}}{{Hartman} et~al.}{2011}]{hartman11}
{Hartman} J.~M.,  {Galloway} D.~K.,    {Chakrabarty} D.,  2011, \apj, 726, 26

\bibitem[\protect\citeauthoryear{{Heinke}, {Rybicki}, {Narayan} \&
  {Grindlay}}{{Heinke} et~al.}{2006}]{heinke06}
{Heinke} C.~O.,  {Rybicki} G.~B.,  {Narayan} R.,    {Grindlay} J.~E.,  2006,
  \apj, 644, 1090

\bibitem[\protect\citeauthoryear{{Hynes}, {Charles}, {Garcia}, {Robinson},
  {Casares}, {Haswell}, {Kong}, {Rupen}, {Fender}, {Wagner}, {Gallo}, {Eves},
  {Shahbaz} \& {Zurita}}{{Hynes} et~al.}{2004}]{hynes04}
{Hynes} R.~I.,  {Charles} P.~A.,  {Garcia} M.~R.,  {Robinson} E.~L.,  {Casares}
  J.,  {Haswell} C.~A.,  {Kong} A.~K.~H.,  {Rupen} M.,  {Fender} R.~P.,
  {Wagner} R.~M.,  {Gallo} E.,  {Eves} B.~A.~C.,  {Shahbaz} T.,    {Zurita} C.,
   2004, \apjl, 611, L125

\bibitem[\protect\citeauthoryear{{King} \& {Ritter}}{{King} \&
  {Ritter}}{1998}]{kingritter98}
{King} A.~R.,  {Ritter} H.,  1998, \mnras, 293, L42

\bibitem[\protect\citeauthoryear{{Kong}, {Kuulkers}, {Charles} \&
  {Homer}}{{Kong} et~al.}{2000}]{kong00}
{Kong} A.~K.~H.,  {Kuulkers} E.,  {Charles} P.~A.,    {Homer} L.,  2000,
  \mnras, 312, L49

\bibitem[\protect\citeauthoryear{{Kong}, {McClintock}, {Garcia}, {Murray} \&
  {Barret}}{{Kong} et~al.}{2002}]{kong02}
{Kong} A.~K.~H.,  {McClintock} J.~E.,  {Garcia} M.~R.,  {Murray} S.~S.,
  {Barret} D.,  2002, \apj, 570, 277

\bibitem[\protect\citeauthoryear{{Kuulkers}, {in't Zand} \&
  {Lasota}}{{Kuulkers} et~al.}{2009}]{kuulkers09}
{Kuulkers} E.,  {in't Zand} J.~J.~M.,    {Lasota} J.,  2009, \aap, 503, 889

\bibitem[\protect\citeauthoryear{{Lasota}}{{Lasota}}{2001}]{lasota01}
{Lasota} J.,  2001, New Astronomy Reviews, 45, 449

\bibitem[\protect\citeauthoryear{{Levine}, {Bradt}, {Cui}, {Jernigan},
  {Morgan}, {Remillard}, {Shirey} \& {Smith}}{{Levine} et~al.}{1996}]{levine96}
{Levine} A.~M.,  {Bradt} H.,  {Cui} W.,  {Jernigan} J.~G.,  {Morgan} E.~H.,
  {Remillard} R.,  {Shirey} R.~E.,    {Smith} D.~A.,  1996, \apjl, 469, L33+

\bibitem[\protect\citeauthoryear{{Li}, {Garcia}, {Forman}, {Jones}, {Kraft},
  {Lal}, {Murray} \& {Wang}}{{Li} et~al.}{2010}]{ligarciaetal10}
{Li} Z.,  {Garcia} M.~R.,  {Forman} W.~R.,  {Jones} C.,  {Kraft} R.~P.,  {Lal}
  D.~V.,  {Murray} S.~S.,    {Wang} Q.~D.,  2010, ApJL, submitted,
  arXiv:1011.1224

\bibitem[\protect\citeauthoryear{{Maitra}, {Markoff} \& {Falcke}}{{Maitra}
  et~al.}{2009}]{maitra09}
{Maitra} D.,  {Markoff} S.,    {Falcke} H.,  2009, \aap, 508, L13

\bibitem[\protect\citeauthoryear{{Menou}, {Hameury}, {Lasota} \&
  {Narayan}}{{Menou} et~al.}{2000}]{menou00}
{Menou} K.,  {Hameury} J.,  {Lasota} J.,    {Narayan} R.,  2000, \mnras, 314,
  498

\bibitem[\protect\citeauthoryear{{Menou} \& {McClintock}}{{Menou} \&
  {McClintock}}{2001}]{menoumcclintock01}
{Menou} K.,  {McClintock} J.~E.,  2001, \apj, 557, 304

\bibitem[\protect\citeauthoryear{{Miller-Jones}, {Gallo}, {Rupen},
  {Mioduszewski}, {Brisken}, {Fender}, {Jonker} \& {Maccarone}}{{Miller-Jones}
  et~al.}{2008}]{millerjones08}
{Miller-Jones} J.~C.~A.,  {Gallo} E.,  {Rupen} M.~P.,  {Mioduszewski} A.~J.,
  {Brisken} W.,  {Fender} R.~P.,  {Jonker} P.~G.,    {Maccarone} T.~J.,  2008,
  \mnras, 388, 1751

\bibitem[\protect\citeauthoryear{{Muno}, {Pfahl}, {Baganoff}, {Brandt}, {Ghez},
  {Lu} \& {Morris}}{{Muno} et~al.}{2005}]{muno05}
{Muno} M.~P.,  {Pfahl} E.,  {Baganoff} F.~K.,  {Brandt} W.~N.,  {Ghez} A.,
  {Lu} J.,    {Morris} M.~R.,  2005, \apjl, 622, L113

\bibitem[\protect\citeauthoryear{{Muno}, {Wijnands}, {Wang}, {Park}, {Brandt},
  {Bauer} \& {Wang}}{{Muno} et~al.}{2007}]{muno07}
{Muno} M.~P.,  {Wijnands} R.,  {Wang} Q.~D.,  {Park} S.,  {Brandt} W.~N.,
  {Bauer} F.~E.,    {Wang} Z.,  2007, The Astronomer's Telegram, 1013

\bibitem[\protect\citeauthoryear{{Papitto}, {Riggio}, {Burderi}, {Di Salvo},
  {D'A{\'{\i}}} \& {Iaria}}{{Papitto} et~al.}{2011}]{papitto11}
{Papitto} A.,  {Riggio} A.,  {Burderi} L.,  {Di Salvo} T.,  {D'A{\'{\i}}} A.,
   {Iaria} R.,  2011, A\&A, in press, arXiv:1006.1303

\bibitem[\protect\citeauthoryear{{Parmar}, {Williams}, {Kuulkers}, {Angelini}
  \& {White}}{{Parmar} et~al.}{1997}]{parmar97}
{Parmar} A.~N.,  {Williams} O.~R.,  {Kuulkers} E.,  {Angelini} L.,    {White}
  N.~E.,  1997, \aap, 319, 855

\bibitem[\protect\citeauthoryear{{Patruno}}{{Patruno}}{2010}]{patruno10}
{Patruno} A.,  2010, \apj, 722, 909

\bibitem[\protect\citeauthoryear{{Porquet}, {Grosso}, {Predehl}, {Hasinger},
  {Yusef-Zadeh}, {Aschenbach}, {Trap}, {Melia}, {Warwick}, {Goldwurm},
  {B{\'e}langer}, {Tanaka}, {Genzel}, {Dodds-Eden}, {Sakano} \&
  {Ferrando}}{{Porquet} et~al.}{2008}]{porquet08}
{Porquet} D.,  {Grosso} N.,  {Predehl} P.,  {Hasinger} G.,  {Yusef-Zadeh} F.,
  {Aschenbach} B.,  {Trap} G.,  {Melia} F.,  {Warwick} R.~S.,  {Goldwurm} A.,
  {B{\'e}langer} G.,  {Tanaka} Y.,  {Genzel} R.,  {Dodds-Eden} K.,  {Sakano}
  M.,    {Ferrando} P.,  2008, \aap, 488, 549

\bibitem[\protect\citeauthoryear{{Porquet}, {Predehl}, {Aschenbach}, {Grosso},
  {Goldwurm}, {Goldoni}, {Warwick} \& {Decourchelle}}{{Porquet}
  et~al.}{2003}]{porquet03}
{Porquet} D.,  {Predehl} P.,  {Aschenbach} B.,  {Grosso} N.,  {Goldwurm} A.,
  {Goldoni} P.,  {Warwick} R.~S.,    {Decourchelle} A.,  2003, \aap, 407, L17

\bibitem[\protect\citeauthoryear{{Rutledge}, {Bildsten}, {Brown}, {Pavlov} \&
  {Zavlin}}{{Rutledge} et~al.}{1999}]{rutledgeetal99}
{Rutledge} R.~E.,  {Bildsten} L.,  {Brown} E.~F.,  {Pavlov} G.~G.,    {Zavlin}
  V.~E.,  1999, \apj, 514, 945

\bibitem[\protect\citeauthoryear{{Rutledge}, {Bildsten}, {Brown}, {Pavlov} \&
  {Zavlin}}{{Rutledge} et~al.}{2000}]{rutledgeetal00}
{Rutledge} R.~E.,  {Bildsten} L.,  {Brown} E.~F.,  {Pavlov} G.~G.,    {Zavlin}
  V.~E.,  2000, \apj, 529, 985

\bibitem[\protect\citeauthoryear{{Rutledge}, {Bildsten}, {Brown}, {Pavlov} \&
  {Zavlin}}{{Rutledge} et~al.}{2001a}]{rutledgeetal01b}
{Rutledge} R.~E.,  {Bildsten} L.,  {Brown} E.~F.,  {Pavlov} G.~G.,    {Zavlin}
  V.~E.,  2001a, \apj, 559, 1054

\bibitem[\protect\citeauthoryear{{Rutledge}, {Bildsten}, {Brown}, {Pavlov} \&
  {Zavlin}}{{Rutledge} et~al.}{2001b}]{rutledgeetal01a}
{Rutledge} R.~E.,  {Bildsten} L.,  {Brown} E.~F.,  {Pavlov} G.~G.,    {Zavlin}
  V.~E.,  2001b, \apj, 551, 921

\bibitem[\protect\citeauthoryear{{Rutledge}, {Bildsten}, {Brown}, {Pavlov} \&
  {Zavlin}}{{Rutledge} et~al.}{2002}]{rutledge02}
{Rutledge} R.~E.,  {Bildsten} L.,  {Brown} E.~F.,  {Pavlov} G.~G.,    {Zavlin}
  V.~E.,  2002, \apj, 577, 346

\bibitem[\protect\citeauthoryear{{Rutledge}, {Bildsten}, {Brown}, {Pavlov},
  {Zavlin} \& {Ushomirsky}}{{Rutledge} et~al.}{2002}]{rutledge_ks1731_02}
{Rutledge} R.~E.,  {Bildsten} L.,  {Brown} E.~F.,  {Pavlov} G.~G.,  {Zavlin}
  V.~E.,    {Ushomirsky} G.,  2002, \apj, 580, 413

\bibitem[\protect\citeauthoryear{{Sakano}, {Warwick}, {Decourchelle} \&
  {Wang}}{{Sakano} et~al.}{2005}]{sakano05}
{Sakano} M.,  {Warwick} R.~S.,  {Decourchelle} A.,    {Wang} Q.~D.,  2005,
  \mnras, 357, 1211

\bibitem[\protect\citeauthoryear{{Shahbaz}, {Dhillon}, {Marsh}, {Casares},
  {Zurita} \& {Charles}}{{Shahbaz} et~al.}{2010}]{shahbaz10}
{Shahbaz} T.,  {Dhillon} V.~S.,  {Marsh} T.~R.,  {Casares} J.,  {Zurita} C.,
  {Charles} P.~A.,  2010, \mnras, 403, 2167

\bibitem[\protect\citeauthoryear{{Tsujimoto}, {Guainazzi}, {Plucinsky},
  {Beardmore}, {Ishida}, {Natalucci}, {Posson-Brown}, {Read}, {Saxton} \&
  {Shaposhnikov}}{{Tsujimoto} et~al.}{2011}]{tsujimoto11}
{Tsujimoto} M.,  {Guainazzi} M.,  {Plucinsky} P.~P.,  {Beardmore} A.~P.,
  {Ishida} M.,  {Natalucci} L.,  {Posson-Brown} J.~L.~L.,  {Read} A.~M.,
  {Saxton} R.~D.,    {Shaposhnikov} N.~V.,  2011, \aap, 525, A25+

\bibitem[\protect\citeauthoryear{{Ushomirsky} \& {Rutledge}}{{Ushomirsky} \&
  {Rutledge}}{2001}]{ushomirskyrutledge01}
{Ushomirsky} G.,  {Rutledge} R.~E.,  2001, \mnras, 325, 1157

\bibitem[\protect\citeauthoryear{{Wagner}, {Starrfield}, {Hjellming}, {Howell}
  \& {Kreidl}}{{Wagner} et~al.}{1994}]{wagner94}
{Wagner} R.~M.,  {Starrfield} S.~G.,  {Hjellming} R.~M.,  {Howell} S.~B.,
  {Kreidl} T.~J.,  1994, \apjl, 429, L25

\bibitem[\protect\citeauthoryear{{Wijnands}, {in't Zand}, {Rupen}, {Maccarone},
  {Homan}, {Cornelisse}, {Fender}, {Grindlay}, {van der Klis}, {Kuulkers},
  {Markwardt}, {Miller-Jones} \& {Wang}}{{Wijnands} et~al.}{2006}]{wijnands06}
{Wijnands} R.,  {in't Zand} J.~J.~M.,  {Rupen} M.,  {Maccarone} T.,  {Homan}
  J.,  {Cornelisse} R.,  {Fender} R.,  {Grindlay} J.,  {van der Klis} M.,
  {Kuulkers} E.,  {Markwardt} C.~B.,  {Miller-Jones} J.~C.~A.,    {Wang} Q.~D.,
   2006, \aap, 449, 1117

\bibitem[\protect\citeauthoryear{{Yuan}, {Quataert} \& {Narayan}}{{Yuan}
  et~al.}{2004}]{yuan04}
{Yuan} F.,  {Quataert} E.,    {Narayan} R.,  2004, \apj, 606, 894

\bibitem[\protect\citeauthoryear{{Zampieri}, {Turolla}, {Zane} \&
  {Treves}}{{Zampieri} et~al.}{1995}]{zampierietal95}
{Zampieri} L.,  {Turolla} R.,  {Zane} S.,    {Treves} A.,  1995, \apj, 439, 849

\bibitem[\protect\citeauthoryear{{Zhao}, {Young}, {Herrnstein}, {Ho},
  {Tsutsumi}, {Lo}, {Goss} \& {Bower}}{{Zhao} et~al.}{2003}]{zhao03}
{Zhao} J.,  {Young} K.~H.,  {Herrnstein} R.~M.,  {Ho} P.~T.~P.,  {Tsutsumi} T.,
   {Lo} K.~Y.,  {Goss} W.~M.,    {Bower} G.~C.,  2003, \apjl, 586, L29

\end{thebibliography}

\clearpage

\end{document}